\documentclass[fleqn,twoside,twocolumn,nofootinbib,showkeys]{revtex4} 
\usepackage[nocpr]{ujp} 

\begin{document}
\title[Generalization of the van der Waals Equation]
{GENERALIZATION\\ OF THE VAN DER WAALS EQUATION\\ FOR ANISOTROPIC FLUIDS IN POROUS MEDIA}%
\author{M.F.~Holovko}
\affiliation{Institute for Condensed Matter Physics of the Nat.
Acad. of Sci. of Ukraine}
\address{1,~Svientsitskii Str., Lviv 79011, Ukraine} 
\email{holovko@icmp.lviv.ua, shmotolokha@icmp.lviv.ua}
\author{V.I.~Shmotolokha}%
\affiliation{Institute for Condensed Matter Physics of the Nat.
Acad. of Sci. of Ukraine}
\address{1,~Svientsitskii Str., Lviv 79011, Ukraine} 
\email{shmotolokha@icmp.lviv.ua} \udk{???} 
\razd{\secvi}

\autorcol{M.F.\hspace*{0.7mm}Holovko,
V.I.\hspace*{0.7mm}Shmotolokha}

\setcounter{page}{769}%

\begin{abstract}
The generalized van der Waals equation of state for anisotropic
liquids in porous media consists of two terms.\,\,One of them is
based on the equation of state for hard spherocylinders in random
porous media obtained from the scaled particle theory.\,\,The second
term is expressed in terms of the mean value of attractive
intermolecular interactions.\,\,The obtained equation is used for
the investigation of the gas-liquid-nematic phase behavior of a
molecular system depending on the anisotropy of molecule shapes,
anisotropy of attractive intermolecular interactions, and porosity
of a porous medium.\,\,It is shown that the anisotropic phase is
formed by the anisotropy of attractive intermolecular interactions
and by the anisotropy of molecular shapes.\,\,The anisotropy of
molecular shapes shifts the phase diagram to lower densities and
higher temperatures.\,\,The anisotropy of attractive interactions
widens significantly the coexistence region between the isotropic
and anisotropic phases and shifts it to the region of lower
densities and higher temperatures.\,\,It is shown that, for
sufficiently long spherocylinders, the liquid-gas transition is
localized completely within the nematic region.\,\,For all the
considered cases, the decrease of the porosity shifts the phase
diagram to the region of lower densities and lower temperatures.
\end{abstract}
 \maketitle

\section{Introduction}\label{sec_hol1}

It is a great pleasure for us to present our paper for publication
in this special issue dedicated to the 70-th anniversary of
Academician L.A. Bulavin, a known Ukrainian scientist in the physics
of liquid state.\,\,His contribution to the development of the
physics of liquid state is important.\,\,Some results obtained by
him are partially summarized in books
\cite{Bul1,Bul2,Bul3,Adam4,Bul5}.\,\,At the same time, the progress
of his former students inside and outside of Ukraine illustrates the
importance of the scientific school created by L.A.~Bulavin.

The first understanding of the nature of the liquid state of matter is
connected with the van der Waals equation formulated nearly 150 years ago
\cite{VanderVaals6}.
This equation provided the possibility to describe the phase
transition from the gaseous to the liquid state and to account for
the presence of the critical point beyond which the gaseous phase
can not be transformed into a liquid.\,\,It also provided the
possibility to describe the coexistence between the liquid and
gaseous phases, and to predict the existence of metastable states,
namely a supercooled gas and a superheated liquid.\,\,The background
of the van der Waals equation is based on the idea of different
treatments of
 short-range repulsive and  long-range attractive intermolecular interactions.\,\,The repulsive interactions fix the size and the
 shape of molecules and essentially determine the structural and entropic properties.\,\,The  contribution of attractive interactions is mainly energetic
 and can be described in the framework of the mean field approximation.\,\,The first strong statistical mechanics treatment of the van der Waals equation was done about 50 years ago for the hard sphere model with attractive interactions in the form of the Kac potential \cite{Lebow66,Kac8}
\begin{equation}
U^{\rm att}(r)=\gamma^{3}U(\gamma r),
\label{Hol1}%
\end{equation}
where $r$ is the interparticle distance.

In the limit $\gamma\rightarrow0,$ the equation of state can be
presented in the form \cite{Kac8,Lebow66,Yukhn9}\vspace*{-2mm}
\begin{equation}
\frac{\beta P}{\rho}=\left(\!\frac{\beta P}{\rho}\!\right)_{\!\!\rm
HS}-12a\eta\beta, \label{Hol2}
\end{equation}
where $\beta=1/(kT)$, $k$ is the Boltzmann constant, $T$ is the
temperature, $P$ is the pressure of the fluid, $\rho$ is the
density, $\left(\!\frac{\beta P}{\rho}\!\right)_{\!\rm HS}$ is the
contribution of hard spheres (HS), $\eta=\frac16\pi\rho D^{3}$ is
the fluid packing fraction, and $D$ is the diameter of a hard
sphere.\,\,The second term in Eq.\,(\ref{Hol2}) describes the
contribution from attractive interactions through the constant
$a$:\vspace*{-2mm}
\begin{equation}
a=-\frac{1}{D^{3}}\int\limits_{D}^{\infty}\gamma^{3}U(\gamma
r)r^{2}{\rm d}r.
\label{Hol3}%
\end{equation}\vspace*{-4mm}

The equation of state in the form (\ref{Hol2}) is a generalization
of the van der Waals equation.\,\,It coincides with it exactly in
the one-dimensional case where the contribution from hard spheres is
described by the Tonks equation \cite{Tonks10}.\,\,However, we can
use a more correct description of the hard sphere contribution such
as, for example, the Carnahan--Starling equation \cite{Carnahan11}.
The equation of state in the form (\ref{Hol2}) can be used also for
the description of non-spherical molecules and can be generalized
for the description of isotropic and anisotropic fluids in porous
media.\,\,This is the aim of this paper.

To this end, we will use the Madden--Glandt model
\cite{Madden12}.\,\,According to this model, a porous medium is
presented as a quenched configuration of randomly distributed hard
spheres forming a matrix, in the free space of which there are fluid
molecules.\,\,A specific description of a fluid in such porous media
is related to double quenched-annealed averages: the annealed
average is taken over all the fluid configurations and an additional
quenched average should be taken over all realizations of the
matrix.

The analytical results for a hard sphere fluid in hard sphere
matrices \cite{Pat15,Hol16} obtained recently by extending the
scaled particle theory (SPT) \cite{Reiss13,Reiss14} provide a strong
basis for a generalization of the van der Waals equation for simple
fluids in porous media \cite{Hol16}.\,\,The generalization of the
results obtained in \cite{Pat15,Hol16} to non-spherical molecules in
porous media \cite{Hol17,HolShmotPat} allowed us to generalize the
van der Waals equation \cite{Hol17}  to anisotropic fluids in porous
media.\,\,The investigations of the gas-liquid-nematic phase
equilibria in the framework of the generalized van der Waals
equation show a rich variety of phase behaviors that depends on the
molecular shape, value of  attractive intermolecular interactions,
and porosity of porous media.\,\,In this paper, we will continue
this investigation.\,\,We will focus on the role of the anisotropy
of attractive interactions.

The paper has the following structure.\,\,In Section~\ref{sec_hol2},
we present the results for a hard spherocylinder fluid in porous
media.\,\,In Section~\ref{sec_hol3}, we use these results to
generalize the van der Waals equation.\,\,In Section~\ref{sec_hol4},
we study the influence of the anisotropy of intermolecular
interactions on the phase behavior of a molecular fluid in a porous
medium, by using the generalized van der Waals equation.

\section{Application of the Scaled\\ Particle Theory to the Description\\ of Thermodynamic Properties\\ of a Spherocylinder
Fluid\\ in a Random Porous Medium}\label{sec_hol2}

A hard spherocylinder fluid is widely used for the description of
the influence of the molecular shape on the orientational ordering
in anisotropic fluids \cite{Onsager19,Vroege20}.\,\,In this section,
we apply the scaled particle theory to the description of the
thermodynamic properties of hard spherocylinders in random porous
media created by hard spheres.\,\,The key point of the SPT theory is
based on the derivation of the chemical potential of an additional
scaled particle of a variable size inserted into a fluid.\,\,This
excess chemical potential is equal to the work needed to create a
cavity in a fluid, which is free from any other particles.\,\,The
theory combines the exact consideration of an infinitely small
particle with the thermodynamic consideration of a scaled particle
of a sufficiently large size.\,\,The exact result for a point scaled
particle in a hard sphere fluid confined in a random porous medium
was obtained in \cite{HolDong21}.\,\,However, this approach named
SPT1 contains a subtle inconsistency appearing, when the size of
matrix particles is quite large compared to the size of fluid
particles.\,\,Later on, this inconsistency was eliminated in a new
approach known as SPT2 \cite{Pat15}.\,\,In this section, we
generalize this approach for hard spherocylinder fluids in random
porous \mbox{media.}\looseness=1

Following \cite{Cotter22,Cotter23,Hol17}, we introduce an additional spherocylinder with the
scaling diameter $D_{s}$ and the scaling length $L_{s}$ into a
spherocylinder fluid in a porous medium:
\begin{equation}
D_{s}=\lambda_{s}D_{1}, \quad L_{s}=\alpha_{s}L_{1},
\label{Hol4}
\end{equation}
where $D_{1}$ and $L_{1}$ are, respectively, the diameter and the
length of the fluid spherocylinder.\,\,The  excess chemical
potential for a small scaled particle can be  written in the form
\cite{Hol17}\vspace*{-3mm}
\[
\beta\mu_{s}^{\rm ex}=-\ln
p_{0}(\alpha_{s},\lambda_{s})-\ln\Biggl[1-\frac{\eta_{1}}{V_{1}p_{0}(\alpha_{s},\lambda_{s})}\,\times\]\vspace*{-7mm}
\[\times\Biggl(\!\frac{\pi}{6}D_{1}^{3}(1+\lambda_{s})^{3}
+\frac{\pi}{4}D_{1}^{2}L_{1}(1+\lambda_{s})^{2}(1+\alpha_{s})\,+\]\vspace*{-5mm}
\[+\,\frac{\pi}{4}D_{1}L_{1}^{2}(1+\lambda_{s})\alpha_{s}\,\times\]\vspace*{-7mm}
\begin{equation}
\times\int\! f(\Omega_{1})f({\Omega_{2}})\sin\vartheta_{12}d
\Omega_{1}d\Omega_{2}\!\Biggr)\!\Biggr]\!, \label{Hol5}
\end{equation}
where $\eta_{1}=\rho_{1}V_{1}$ is the fluid packing fraction, $\rho_{1}$
is the fluid density, $V_{1}$ is the spherocylinder volume, and
\begin{equation}
p_{0}(\alpha_{s},\lambda_{s})=\exp[-\beta\mu_{s}^{0}(\alpha_{s},\lambda_{s})]
\label{Hol6}
\end{equation}
is the probability to find a cavity created by a scaled particle in
the empty matrix.\,\,It is defined by the excess chemical potential
$\mu_{s}^{0}(\alpha_{s},\lambda_{s})$ of the scaled particle in the
limit of an infinite dilution, $\Omega=(\vartheta,\varphi)$ denotes
the orientation of particles and is defined by the angles
$\vartheta$ and $\varphi$; $d\Omega=\frac{1}{4\pi}\sin\vartheta
d\vartheta d\varphi$ is the normalized angle element,
$\vartheta_{12}$ is the angle between orientational vectors of two
molecules, and $f(\Omega)$ is the singlet orientational distribution
function normalized in such a way that
\begin{equation}
\int f (\Omega)d\Omega=1.
\label{Hol7}
\end{equation}

We note that, here and below, we use conventional notations
\cite{Pat15,Hol16,Hol17,HolShmotPat}, where the index ``1'' is used
to denote a fluid component, the index ``0'' denotes matrix
particles, while, for the scaled particles, the index ``s'' is used.

For a large scaled particle, the excess chemical potential is given by the
thermodynamic expression for the work needed to create a macroscopic cavity
inside a fluid and can be presented in the form
\begin{equation}
\beta\mu_{s}^{\rm ex}=w(\alpha_{s},\lambda_{s})+{\beta PV_{s}} /{
p_{0}(\lambda_{s},\alpha_{s})}, \label{Hol8}
\end{equation}
where $P$ is the pressure of the fluid, and $V_{s}$ is the volume of
the scaled particle.\,\,The multiplier
$1/p_{0}(\alpha_{s},\lambda_{s})$ appears due to an excluded volume
occupied by matrix particles.\,\,The  probability
$p_{0}(\alpha_{s},\lambda_{s})$ is directly related to two different
types of porosity introduced by us in
\cite{Pat15,Hol16,Hol17,HolShmotPat}.\,\,The first one corresponds
to the geometrical porosity
\begin{eqnarray}
\phi_{0}=p_{0}\left(\alpha_{s}=\lambda_{s}=0 \right)\!, \label{Hol9}
\end{eqnarray}
characterizing the free volume for a fluid.\,\,For a hard sphere
matrix,
\begin{equation}
\phi_{0}=1-\eta_{0},
\label{Hol10}
\end{equation}
where $\eta_{0}=\frac16\pi D^{3}_{0}\rho_{0}$ is the packing fraction of the
matrix, $\rho_{0}$ is the density of matrix particles, and $D_{0}$ is the
diameter of matrix particles.

The second type of porosity corresponds to the case
$\lambda_{s}=\alpha_{s}=1$ and leads to the thermodynamic porosity
\begin{eqnarray}
\phi=p_{0}(\alpha_{s}=\lambda_{s}=1)=\exp(-\beta\mu_{1}^{0})
\label{Hol11}
\end{eqnarray}
defined by the excess chemical potential of fluid particles $\mu_{1}^{0}$ in
the limit of infinite dilution. It characterizes the adsorption of a fluid in
the empty matrix. In the case under consideration,
\[
\phi=(1-\eta_{0})\exp\Biggl[-\frac{\eta_{0}}{1-\eta_{0}}\tau\left(\!\frac
{3}{2}(\gamma_{1}+1)+3\gamma_{1}\tau\!\right)-\]\vspace*{-7mm}
\[-\,\frac{\eta_{0}^{2}}{(1-\eta_{0})^{2}}\frac{9}{2}\gamma_{1}\tau^{2}\,-\]\vspace*{-7mm}
\begin{equation}
-\,\frac{\eta_{0}}{(1-\eta_{0})^{3}}(3\gamma_{1}-1)\frac12\tau^{3}(1+\eta_{0}+\eta_{0}^{2})\!\Biggr]\!,
\label{Hol12}
\end{equation}
where $\tau=\frac{D_{1}}{D_{0}}$, and $\gamma_{1}=1+L_{1}/D_{1}$.

In accordance with the ansatz of the SPT theory
\cite{Reiss13,Reiss14,Pat15,Hol16,Hol17,HolShmotPat},
$w(\lambda_{s},\alpha_{s})$ can be presented in the form
\[
w(\lambda_{s},\alpha_{s})=w_{{00}}+w_{{10}}
\lambda_{s}\,+\]\vspace*{-9mm}
\begin{equation}
+\,w_{{01}}\alpha_{s}+w_{{11}} \alpha_{s}\lambda_{s}+\frac
{w_{{20}}\lambda_{s}^{2}}{2}. \label{Hol13}
\end{equation}
The coefficients of this expansion can be found from the continuity
of $\mu_{s}^{\rm ex}$ and the corresponding derivatives
$\partial\mu_{s}^{\rm ex}/\partial\lambda_{s}$,
$\partial\mu_{s}^{\rm ex}/\partial\alpha_{s}$,
$\partial^{2}\mu_{s}^{\rm
ex}/(\partial\alpha_{s})(\partial\lambda_{s})$ and
$\partial^{2}\mu_{s}^{\rm ex}/\partial\lambda_{s}^{2}$ at
$\lambda_{s}=\alpha_{s}=0$.\,\,After setting
$\lambda_{s}=\alpha_{s}=1,$ we found the relation between the
pressure $P$ and the excess chemical potential $\mu_{1}^{\rm ex}$ of
a fluid:
\[
\beta\left(\mu_{1}^{\rm ex}-\mu_{1}^{0}\right)=-\ln
\left(1-\eta_{1}/\phi_{0}\right)+A(\tau(f))\,\times\]\vspace*{-9mm}
\begin{equation}
\times\,\frac{\eta_{1}/\phi_{0}}{1-\eta_{1}/\phi_{0}}+B(\tau(f))\frac{(\eta_{1}/\phi_{0})^{2}}{(1-\eta_{1}/\phi_{0})^{2}}
+\frac{\beta P}{\rho_{1}}\frac{\eta_{1}}{\phi}, \label{Hol14}
\end{equation}
where\vspace*{-3mm}
\[
A(\tau(f))
=6+\frac{6\left(\gamma_{1}-1\right)^2\tau(f)}{3\gamma_{1}-1}\,-\]\vspace*{-7mm}
\[-\,\frac{p'_{0\lambda}}{\phi_0}\left(\!4+\frac{3\left(\gamma_{1}-\!\right)^2\tau(f)}{3\gamma_{1}-1}\right)-\frac{p'_{0\alpha}}{\phi_0}\left(\!1+\frac{6\gamma_{1}}{3\gamma_{1}-1}\!\right)-
\]\vspace*{-7mm}
\begin{equation}
-\,\frac{p''_{0\alpha\lambda}}{\phi_0}-\frac{1}{2}
    \frac{p''_{0\lambda\lambda}}{\phi_0}+2\frac{p'_{0\alpha}p'_{0\lambda}}{\phi_0^{2}}+\left(\!\frac{p'_{0\lambda}}{\phi_0}\!\right)^{\!\!2}\!\!,
\label{Hol15}
\end{equation}\vspace*{-7mm}
 \[B(\tau(f)) =
 \left(\!\frac{6\gamma_{1}}{3\gamma_{1}-1}-\frac{p'_{0\lambda}}{\phi_0}\!\right)\!\!\Bigg(\!\frac{3\left(2\gamma_{1}-1\!\right)}{3\gamma_{1}-1}\,+\]\vspace*{-7mm}
\begin{equation}
-\,\frac{3\left(\!
\gamma_{1}-1\right)^2\tau(f)}{3\gamma_{1}-1}\frac{p'_{0\alpha}}{\phi_0}
-\frac{1}{2}\frac{p'_{0\lambda}}{\phi_0}\!\Bigg)\!, \label{Hol16}
 \end{equation}\vspace*{-7mm}
%
\begin{equation}
\tau(f)=\frac{4}{\pi}\int f(\Omega_{1}) f(\Omega_{2}) \sin
\vartheta_{12}d\Omega_{1} d\Omega_{2}. \label{Hol17}
\end{equation}\vspace*{-3mm}

\noindent $ p'_{0\lambda } = \frac{\partial
{p_0}(\alpha_s,\lambda_s)}{\partial \lambda_s}$,
 $p'_{0\alpha } =  \frac{\partial p_{0}(\alpha_s ,\lambda_s)}{\partial \alpha_s}$,
 $ p''_{0\alpha\lambda } =  \frac{\partial^{2} {p_0}(\alpha_s,\lambda_s)}{\partial\alpha_s\partial
 \lambda_s}$,
 $p''_{0\lambda\lambda } = \frac{\partial^2{p_0}(\alpha_s,\lambda_s)}{\partial
\lambda_{s}^{2}}$ are the corresponding derivatives at
$\alpha_{s}=\lambda{s}=0$.

Using the Gibbs--Duhem equation, which relates the pressure $P$ of a
fluid to its total chemical potential
$\mu_{1}=\ln(\Lambda_{1}^{3}\Lambda_{1R})+\mu_{1}^{\rm ex}$,
\begin{equation}
\left(\!\frac{\partial
P}{\partial\rho_{1}}\!\right)_{\!\!T}=\rho_{1}\left(\!\frac{\partial
\mu_{1}}{\partial\rho_{1}}\!\right)_{\!\!T}\!, \label{Hol18}
\end{equation}
one derives the fluid compressibility in the form
\[
\beta\left(\!\frac{\partial
P}{\partial\rho_{1}}\!\right)_{\!\!T}=\frac{1}{\left(1-\eta_{1}/\phi\right)}\,+\]\vspace*{-7mm}
\[+\,(1+A(\tau(f)))\frac{\eta_{1}/\phi_{0}}
{\left({1-\eta_{1}/\phi}\right)
\left(1-\eta_{1}/\phi_{0}\right)}\,+\]\vspace*{-7mm}
\[+\,(A(\tau(f))+2B(\tau(f)))\frac{\left(\eta_{1}/\phi_{0}\right)^{2}}{\left(1-\eta_{1}/\phi\right)\left(1-\eta_{1}/\phi_{0}\right)^{2}}\,+\]\vspace*{-7mm}
\begin{equation}
+\,
2B(\tau(f))\frac{\left(\eta_{1}/\phi_{0}\right)^{3}}{\left(1-\eta_{1}/\phi\right)\left(1-\eta_{1}/\phi_{0}\right)^{3}},
\label{Hol19}
\end{equation}
where $\Lambda_{1}$ is the fluid thermal wave length, and
$\Lambda_{1R}^{-1}$ is the rotational partition function of a single
molecule~\cite{Gray24}.

After the integration of relation (\ref{Hol19}) over $\rho_{1},$ one obtains the
expressions for the chemical potential and for the pressure in the SPT2
approximation \cite{Pat15,Hol16,Hol17}:
\[
\beta(\mu_{1}^{\rm
ex}-\mu_{1}^{0})=-\ln(1-\eta_{1}/\phi)+(A(\tau(f))+1)\,\times\]\vspace*{-9mm}
 \[\times\,\frac{\phi}{\phi-\phi_{0}}\ln\frac{1-\eta_{1}/\phi}{1-\eta_{1}/\phi_{0}}\,+\]\vspace*{-7mm}
\[+\,(A(\tau(f))+2B(\tau(f)))\frac{\phi}{\phi-\phi_{0}}\left(\!\frac{\eta_{1}/\phi_{0}}{1-\eta_{1}/\phi_{0}}\right.\,-\]\vspace*{-7mm}
\[\left.-\, \frac{\phi}{\phi-\phi_{0}}\ln\frac{1-\eta_{1}/\phi}
{1-\eta_{1}/\phi_{0}}\!\right)+2B(\tau(f))\,\times\]\vspace*{-7mm}
\[\times\,\frac{\phi}{\phi-\phi_{0}}\left[\frac12\frac{(\eta_{1}/\phi_{0})^{2}}{(1-\eta_{1}/\phi_{0})^{2}}-\frac{\phi}{\phi-\phi_{0}}
\frac{\eta_{1}/\phi_{0}}{1-\eta_{1}/\phi_{0}}\right.\,+\]\vspace*{-7mm}
\begin{equation}
\left.+\,\frac{\phi^{2}}{(\phi-\phi_{0})^{2}}
\ln\frac{1-\eta_{1}/\phi} {1-\eta_{1}/\phi_{0}}\right]\!\!,
\label{Hol20}
\end{equation}\vspace*{-7mm}
\[
\frac{\beta
P}{\rho_{1}}=-\frac{\phi}{\eta_{1}}\ln\frac{1-\eta_{1}/\phi}{1-\eta_{1}/\phi_{0}}+
(1+A(\tau(f)))\,\times\]\vspace*{-7mm}
\[\times\,\frac{\phi}{\eta_{1}}\frac{\phi}{\phi-\phi_{0}}\ln\frac{1-\eta_{1}/\phi}{1-\eta_{1}/\phi_{0}}\,+\]\vspace*{-7mm}
\[+\,(A(\tau(f))+2B(\tau(f)))\frac{\phi}{\phi-\phi_{0}}\left[\frac{1}{1-\eta_{1}/\phi_{0}}\right.\,-\]\vspace*{-7mm}
\[\left.-\,\frac{\phi}{\eta_{1}}\frac{\phi}{\phi-\phi_{0}}
\ln\frac{1-\eta_{1}/\phi}
{1-\eta_{1}/\phi_{0}}\right]+\]\vspace*{-7mm}
\[+\,2B(\tau(f))\frac{\phi}{\phi-\phi_{0}}\left[\frac12\frac{\eta_{1}/\phi_{0}}{(1-\eta_{1}/\phi_{0})^{2}}-\frac{2\phi-\phi_{0}}
{\phi-\phi_{0}} \,\times\right.\]\vspace*{-7mm}
\begin{equation}
\left.\times\,\frac{1}{1-\eta_{1}/\phi_{0}}+
\frac{\phi}{\eta_{1}}\frac{\phi^{2}}{(\phi-\phi_{0})^{2}}\ln\frac{1-\eta_{1}/\phi}{1-\eta_{1}/\phi_{0}}\right]\!\!,
\label{Hol21}
\end{equation}
where\vspace*{-3mm}
\begin{equation}
\sigma(f)=\int f(\Omega)\ln f(\Omega)d\Omega.
\label{Hol22}
\end{equation}
As noted in \cite{Pat15,Hol16,Hol17,HolShmotPat}, expressions
(\ref{Hol20})-(\ref{Hol21}) have divergences at $\eta_{1}=\phi$ and
$\eta_{1}=\phi_{0}$.\,\,Since $\phi<\phi_{0}$, the divergence at
$\eta_{1}=\phi$ occurs at lower densities and should be removed.
Different corrections improving the SPT2 results were proposed in
\cite{Vroege20,HolDong21,Cotter22}.\,\,In this paper, we consider
only the SPT2b approach, which can be derived if
 $\phi$ is replaced by $\phi_{0}$ everywhere in (\ref{Hol19}) except the first
 term.\,\,In consequence, the chemical potential and the pressure of the fluid can be
 presented in the form
\[
\beta(\mu_{1}^{\rm ex}-\mu_{1}^{0})^{\rm
SPT2b}=\sigma(f)-\ln(1-\eta_{1}/\phi)\,+\]\vspace*{-9mm}
\[+\,(1+A(\tau(f)) )\frac{\eta_{1}/\phi_{0}}{1-\eta_{1}/\phi_{0}}\,+\]\vspace*{-7mm}
\[+\,\frac12(A(\tau(f)) +2B(\tau(f))
)\frac{(\eta_{1}/\phi_{0})^{2}}{(1-\eta_{1}/\phi_{0})^{2}}\,+
\]\vspace*{-7mm}
\begin{equation}
+\,\frac23 B(\tau(f))
\frac{(\eta_{1}/\phi_{0})^{3}}{(1-\eta_{1}/\phi_{0})^{3}},\label{Hol23}
\end{equation}\vspace*{-7mm}
%
\[
\left(\!\frac{\beta P}{\rho_{1}}\!\right)^{\!\!\rm
SPT2b}=-\frac{\phi}{\eta_{1}}\ln\left(\!1-\frac{\eta_{1}}{\phi}\!\right)
+\frac{\phi_{0}}{\eta_{1}}
\ln\left(\!1-\frac{\eta_{1}}{\phi_{0}}\!\right)+\]\vspace*{-7mm}
\[+\,\frac{1}{1-\eta_{1}/{\phi_{0}}}
+\frac{A(\tau(f))}{2}\frac{\eta_{1}/\phi_{0}}{(1-\eta_{1}/\phi_{0})^{2}}\,+\]\vspace*{-7mm}
\begin{equation}
+\,\frac{2B(\tau(f))}{3}\frac{(\eta_{1}/\phi_{0})^{2}}{(1-\eta_{1}/\phi_{0})^{3}},
\label{Hol24}
\end{equation}
which reproduces quite well the results of computer simulations
\cite{Pat15,Hol16,Hol17,HolShmotPat,Hol25}.

From the thermodynamic relation
\begin{equation}
\frac{\beta F}{V}=\beta \mu_1\rho_1-\beta P,
\label{Hol25}
\end{equation}
one can obtain the expression for the free energy:
\[
\beta V^{-1}F^{\rm SPT2b} = \rho_1
\sigma(f)+\rho_1(\ln(\Lambda_1^{3}\rho_1)-1)\,+\]\vspace*{-9mm}
\[+\,\beta\mu_1^{0}\rho_{1}
-\rho_{1}\ln(1- \eta_1/\phi)\,+\]\vspace*{-9mm}
\[+\,\frac{\rho_1\phi}{\eta_1}\ln(1-\eta_1/\phi)
-
\frac{\rho_1\phi_{0}}{\eta_1}\ln(1-\eta_1/\phi_{0})\,+\]\vspace*{-7mm}
\[+\,\rho_1 \frac{A(\tau(f))}{2}
\frac{\eta_1/\phi_0}{1-\eta_1/\phi_0}\,+\]\vspace*{-7mm}
\begin{equation}
+\,\rho_1 \frac{B(\tau(f))}{3}
\left(\!\frac{\eta_1/\phi_0}{1-\eta_1/\phi_0}\!\right)^{\!\!2}\!\!.
\label{Hol26}
\end{equation}

By minimizing the free energy with respect to the variation of the
distribution function $f(\Omega),$ we obtain the integral equation
\begin{equation}
\ln f(\Omega_{1})+1+C\int f(\Omega_{2})\sin\vartheta_{12}d\Omega_{2}=0,
\label{Hol27}
\end{equation}
where\vspace*{-3mm}
\[
C=\frac{\eta_{1}/\phi_{0}}{1-{\eta_{1}/\phi_{0}}}\left[\frac{3(\gamma_{1}-1)^{2}}{3\gamma_{1}-1}\left(\!1-\frac{p'_{0\lambda}}{2\phi_{0}}\!\right)+\right.
\]\vspace*{-7mm}
\begin{equation}
\left.+\,\frac{{\eta_{1}/\phi_{0}}}{(1-{\eta_{1}/\phi_{0}})}\frac{(\gamma_{1}-1)^{2}}{3\gamma_{1}-1}\left(\!\frac{6\gamma_{1}}{3\gamma_{1}-1}
-\frac{p'_{0\lambda}}{\phi_{0}}\!\right)\!\right]\!. \label{Hol28}
\end{equation}

This equation can be solved numerically, by using an iteration
procedure according to the algorithm proposed in
\cite{Herzfeld26}.\,\,We note that, in the Onsager limit
$L_{1}\rightarrow\infty$, $D_{1}\rightarrow0$ \cite{Onsager19},
\begin{equation}
C\rightarrow c=\frac14\pi L_{1}^{2}D_{1}\rho_{1},
\label{Hol29}
\end{equation}
where $c$ is finite.

From the bifurcation analysis, it is found that Eq.\,(\ref{Hol27})
has two characteristic points \cite{Kayser27}:
\begin{equation}
c_{i}=3.290, \quad c_{n}=4.191,
\label{Hol30}
\end{equation}
where $c_{i}$ corresponds to the highest density of a stable
isotropic fluid, and $c_{n}$ is related to the minimum density of a
stable nematic fluid.

In the presence of a porous medium within the Onsager model, we have
\begin{equation}
c_{i}/\phi_{0}=3.290, \quad c_{n}/\phi_{0}=4.191.
\label{Hol31}
\end{equation}
For finite $L_{1}$ and $D_{1},$
\begin{equation}
C_{i}=3.290, \quad C_{n}=4.191,
\label{Hol32}
\end{equation}
where $C_{i}$ and $C_{n}$ are defined by~(\ref{Hol28}).\,\,These
values of $C_{i}$ and $C_{n}$ define the isotropic-nematic phase
diagram depending on the ratio $L_{1}/D_{1}$ and the matrix
parameters for a hard spherocylinder fluid in a matrix.\,\,As was
shown in \cite{Hol17}, the obtained theoretical results  are in
agreement with the data of computer simulations \cite{Schmidt28}.

\section{Generalization of the van der Waals Equation for Anisotropic Fluids in Random Porous Media}\label{sec_hol3}

We will use the results  for a hard spherocylinder fluid presented
in the previous section as a reference system for the generalization
of the van der Waals equation for anisotropic fluids in random
porous media.

Such generalization includes the non-spherical shape of molecules,
anisotropy of the intermolecular interaction, and presence of a
porous medium.\,\,As a result, we will have a more general form of
Eq.\,\,(\ref{Hol2}) \cite{Hol17}:
\begin{equation}
\frac{\beta P}{\rho_{1}}=\left(\!\frac{\beta
P}{\rho_{1}}\!\right)_{\!\!\rm HSC}-12\eta_{1}a\beta, \label{Hol33}
\end{equation}
 where $\left(\!\frac{\beta P}{\rho_{1}}\!\right)_{\!\rm
HSC}$ is the contribution from hard spherocylinders (HSC) in porous
media, which is described by expression (\ref{Hol24}).\,\,The
contribution from attractive interactions is described by a constant
$a,$ which can be presented in the form
 \begin{equation}
a=-\frac{1}{\phi_{0}V_{1}}\int\! f(\Omega_{1})f(\Omega_{2}) U^{\rm
att}(r_{12}\Omega_{1}\Omega_{2})d\bar{r}_{12}d\Omega_{1}d\Omega_{2},
\label{Hol34}
\end{equation}
where the factor $\frac{1}{\phi_{0}}$ excludes the volume occupied
by matrix particles, $V_{1}$ is the volume of a molecule,
$\eta_{1}=\rho_{1}V_{1}$, and $U^{\rm attr}(r_{12}\Omega_{1}\Omega_{2})$
is the attractive part of the intermolecular interaction.

Similarly to \cite{Hol17}, we present the potential $U^{\rm
attr}(r_{12}\Omega_{1}\Omega_{2})$ in the form of a modified
Lennard-Jones potential
\[
U^{\rm att}(r_{12}\Omega_{1}\Omega_{2})=U_{\rm
LJ}\left(\!\frac{\sigma(\Omega_{1}\Omega_{2}\Omega_{r})}{r_{12}}\!\right)\times\]\vspace*{-7mm}
\begin{equation}
\times\left[1+\chi P_{2}(\cos\vartheta_{12})\right], \label{Hol35}
\end{equation}
\begin{equation}
\begin{array}{l}
\displaystyle U_{\rm
LJ}\left(\!\frac{\sigma(\Omega_{1}\Omega_{2}\Omega_{r})}{r_{12}}\!\right)=
4\epsilon_{0}\left[\!\left(\!\frac{\sigma(\Omega_{1}\Omega_{2}\Omega_{r})}{r_{12}}\!\right)^{\!\!12}\right.-   \\
\displaystyle \left.-
\left(\!\frac{\sigma(\Omega_{1}\Omega_{2}\Omega_{r})}{r_{12}}\!\right)^{\!\!6}\right]\!\!,\quad
r_{12}\geq  \sigma(\Omega_{1}\Omega_{2}\Omega_{r}), \\
\displaystyle U_{\rm
LJ}\left(\!\frac{\sigma(\Omega_{1}\Omega_{2}\Omega_{r})}{r_{12}}\!\right)=0,\quad
 r_{12}< \sigma(\Omega_{1}\Omega_{2}\Omega_{r}), \\
\end{array}
\label{Hol36}
\end{equation}
where $P_{2}(\cos\vartheta_{12})=\frac12(3\cos^{2}\vartheta_{12}-1)$
is the second Legendre polynomial, $\vartheta_{12}$ is the angle
between the principal axes of two interacting molecules, and
$\sigma(\Omega_{1}\Omega_{2}\Omega_{r})$ is the contact distance
between molecules.\,\,It depends on the orientations of two
molecules, $\Omega_{1}$ and $\Omega_{2},$ as well as on the
orientation of the distance vector $\bar{r}_{12}$ between their
centers.\,\,In the case where the repulsive part of the interaction
is spherical ($D_{1}=\sigma$), this potential reduces
 to the Maier--Saupe potential \cite{Schmidt28}.\,\,We note that potential (\ref{Hol36}) is the sum of two Lennard-Jones potentials.\,\,The
 first one is related to the isotropic attraction, and the second one to the anisotropic attraction.\,\,The ratio of the well
 depths of these two potentials $\chi=\epsilon_{2}/\epsilon_{0}$ specifies the
 degree of anisotropy in the attraction of the total potential.

Following the traditional scheme \cite{Franko30}, taking into account that
$d\bar{r}=r^{2}drd\Omega_{r}$, and using a dimensionless  intermolecular
distance $r^{*}=r/\sigma(\Omega_{1}\Omega_{2}\Omega_{r})$, one obtains
 \[
a=-\frac{1}{\phi_{0}V_{1}}\int d \Omega_{1}d\Omega_{2}
f(\Omega_{1})f(\Omega_{2})\left[1\!+\!\chi
P_{2}(\cos\vartheta_{12})\right]\times\]\vspace*{-7mm}
\begin{equation}
\times\, V_{1}^{\rm
exc}(\Omega_{1}\Omega_{2})3\int\limits_{0}^{\infty}r^{*2}dr^{*}\beta
U_{\rm LJ}(r^{*}), \label{Hol37}
\end{equation}
where\vspace*{-3mm}
\begin{equation}
V_{\rm exc}(\Omega_{1}\Omega_{2})=\frac13\int
d\Omega_{r}[\sigma(\Omega_{1}\Omega_{2}\Omega_{r})]^{3}
\label{Hol38}
\end{equation}
is the excluded volume formed by two spherocylinders with orientations
$\Omega_{1}$ and $\Omega_{2}$.

The following expressions for the chemical potential and the free
energy correspond to Eq.\,(\ref{Hol33}):
\[
\beta(\mu_{1}^{\rm
ex}-\mu_{1}^{0})=\sigma(f)-\ln(1-\eta_{1}/\phi)\,+\]\vspace*{-9mm}
\[+\,(1+A(\tau(f)) )\frac{\eta_{1}/\phi_{0}}{1-\eta_{1}/\phi_{0}}\,+\]\vspace*{-7mm}
\[+\,\frac12(A(\tau(f))
+2B(\tau(f)))\frac{(\eta_{1}/\phi_{0})^{2}}{(1-\eta_{1}/\phi_{0})^{2}}\,+
\]\vspace*{-7mm}
\begin{equation}
+\,\frac23 B(\tau(f))
\frac{(\eta_{1}/\phi_{0})^{3}}{(1-\eta_{1}/\phi_{0})^{3}}-24\beta\eta_{1}a
\label{Hol39}
\end{equation}
\[\frac{\beta F}{V} = \rho_1
\sigma(f)+\rho_1(\ln(\Lambda_1^{3}\rho_1)-1)+\beta\mu_1^{0}\rho_{1}\,-\]\vspace*{-9mm}
\[-\,\rho_{1}\ln(1-
\eta_1/\phi)+\frac{\rho_1\phi}{\eta_1}\ln(1-\eta_1/\phi)\,-\]\vspace*{-7mm}
\[-\, \frac{\rho_1\phi_{0}}{\eta_1}\ln(1-\eta_1/\phi_{0})+\rho_1 \frac{A(\tau(f))}{2}
\frac{\eta_1/\phi_0}{1-\eta_1/\phi_0}\,+\]\vspace*{-7mm}
\begin{equation}
+\,\rho_1
\frac{B(\tau(f))}{3}\left(\!\frac{\eta_1/\phi_0}{1-\eta_1/\phi_0}\!\right)^{\!\!2}-12\rho_{1}a\eta_{1}\beta.
\label{Hol40}
\end{equation}
From the last expression, we have the following integral equation for the
singlet distribution function:
\[\ln f(\Omega_{1})+1+C\int
f(\Omega_{2})\sin\vartheta_{12}d\Omega_{2}\,+\]\vspace*{-7mm}
\[+\int
f(\Omega_{2})P_{2}(\cos\vartheta_{12})d\Omega_{2}\,\times\]\vspace*{-7mm}
\begin{equation}
\times\,\frac{\beta\rho_{1}\chi}{\phi_{0}}\int U_{\rm
LJ}\left(\!\frac{\sigma(\Omega_{1}\Omega_{2}\Omega_{r})}{r}\!\right)d\bar{r}=0,
\label{Hol41}
\end{equation}
where $C$ is given by expression (\ref{Hol28}).

The obtained singlet distribution function $f(\Omega)$ is used in
(\ref{Hol34}) for the calculation of the parameter $a$.

\section{Influence of Anisotropic\\ Attractive Intermolecular Interactions\\ on the Phase Behavior of Anisotropic\\ Fluids in Porous Media}\label{sec_hol4}

Now, we apply the developed theory to the description of the
gas-liquid-nematic phase behavior of the considered molecular fluids
in porous media created by a random configuration of hard
spheres.\,\,Given the chemical potential and the pressure as
functions of the density $\rho_{1}$ at different temperatures, one
can calculate the coexistence curves from the conditions of
thermodynamic equilibrium:
\begin{equation}
\begin{array}{l}
\mu_{1}(\rho_{1}^{1},T)=\mu_{1}(\rho_{1}^{2}, T),\\[2mm]
P(\rho_{1}^{1},T)=P(\rho_{1}^{2},T), \label{Hol43}
\end{array}
\end{equation}
where $\rho_{1}^{1}$ and $\rho_{1}^{2}$ are the fluid densities of
two different phases 1 and 2.\,\,The numerical solution of these
equations is realized with the use of the Newton--Raphson algorithm.

\begin{figure}%
\vskip1mm
\includegraphics[width=\column]{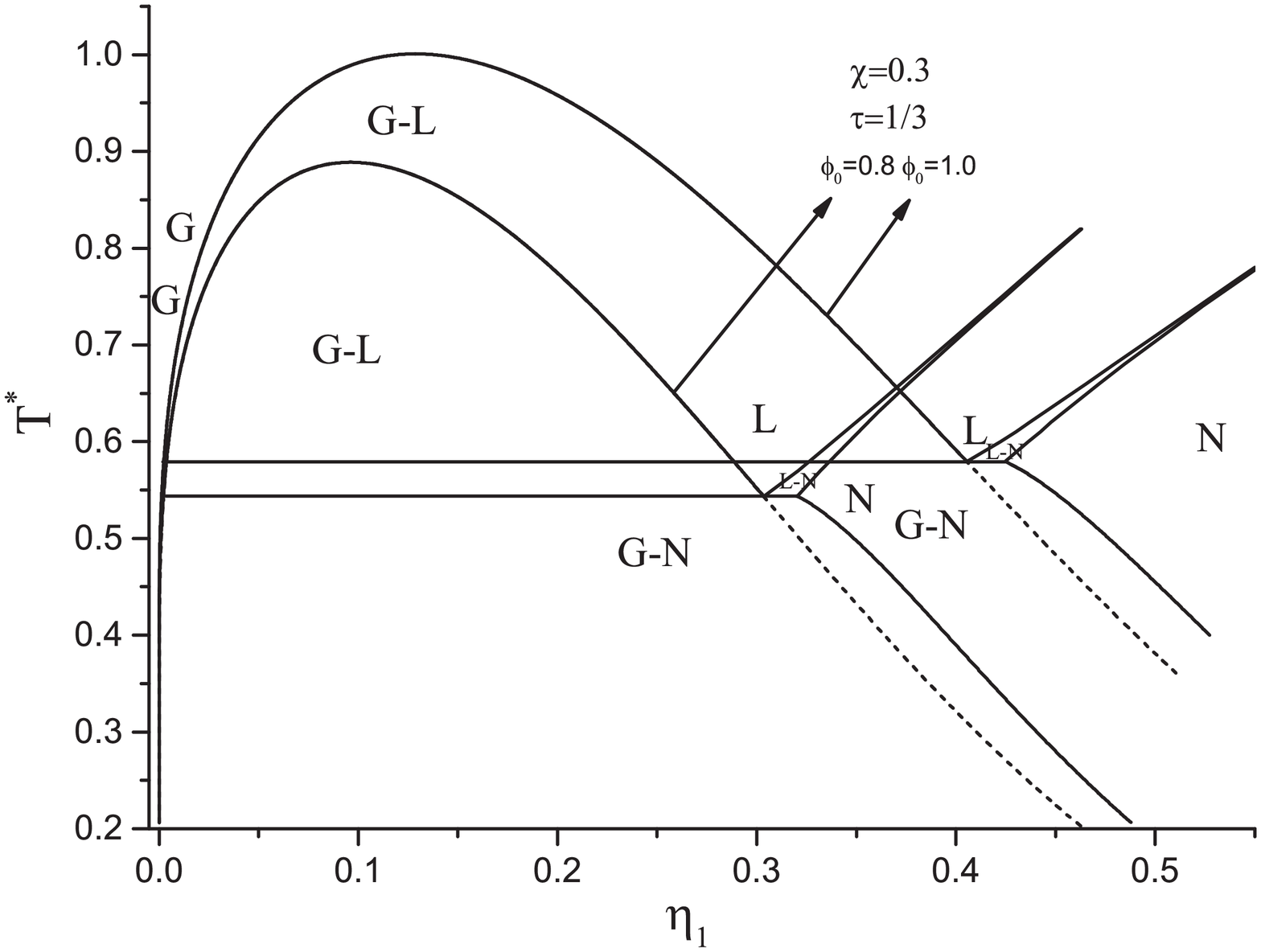}\\
{\it a}\\[3mm]
\includegraphics[width=\column]{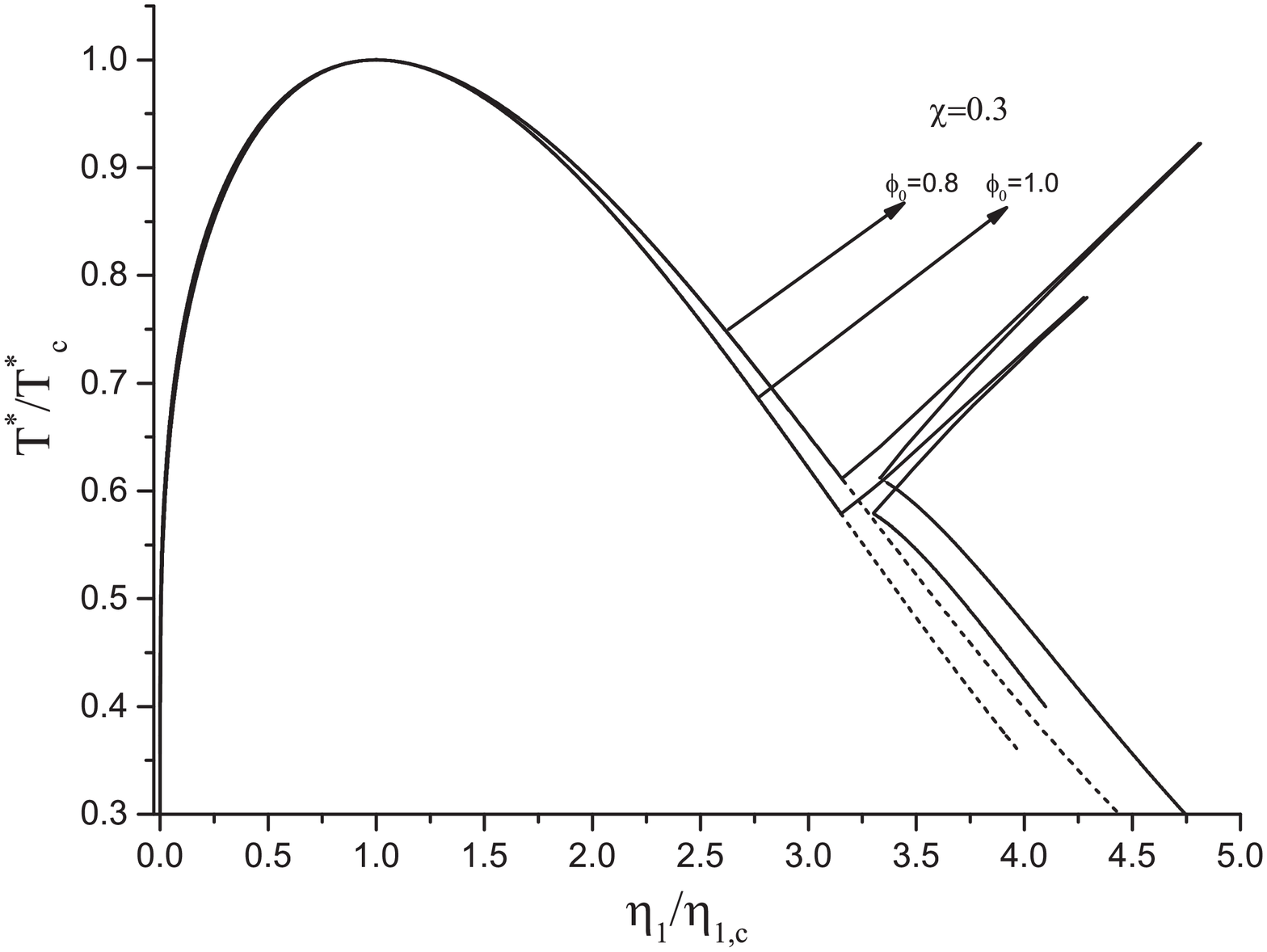}
{\it b} \vskip-3mm\caption{Temperature-density phase diagram
calculated from the generalized van der Waals equation for hard
spheres with isotropic and anisotropic attractive interactions
($\chi=0.3$) in random porous media with the porosity $\phi_{0}=0.8$
and $\tau=$ $=1/3$.\,\,For the purpose of comparison, the phase
diagram in the absence of a porous medium $\phi_{0}=1.0$ is also
presented.\,\,The dashed lines correspond to the case
$\chi=0$.\,\,On the top (case $a$), the phase diagram in the
coordinates
 $T^{*}=kT/\epsilon_{0}$ and $\eta_{1}=\rho_{1}V_{1}$ is presented.\,\,At the bottom (case $b$),
 the phase diagram is presented in the coordinates $T^{*}/T^{*}_{c}$ and $\eta_{1}/\eta_{1,c}$  }\label{Fig1}
\end{figure}

\begin{figure}%
\vskip1mm
\includegraphics[width=\column]{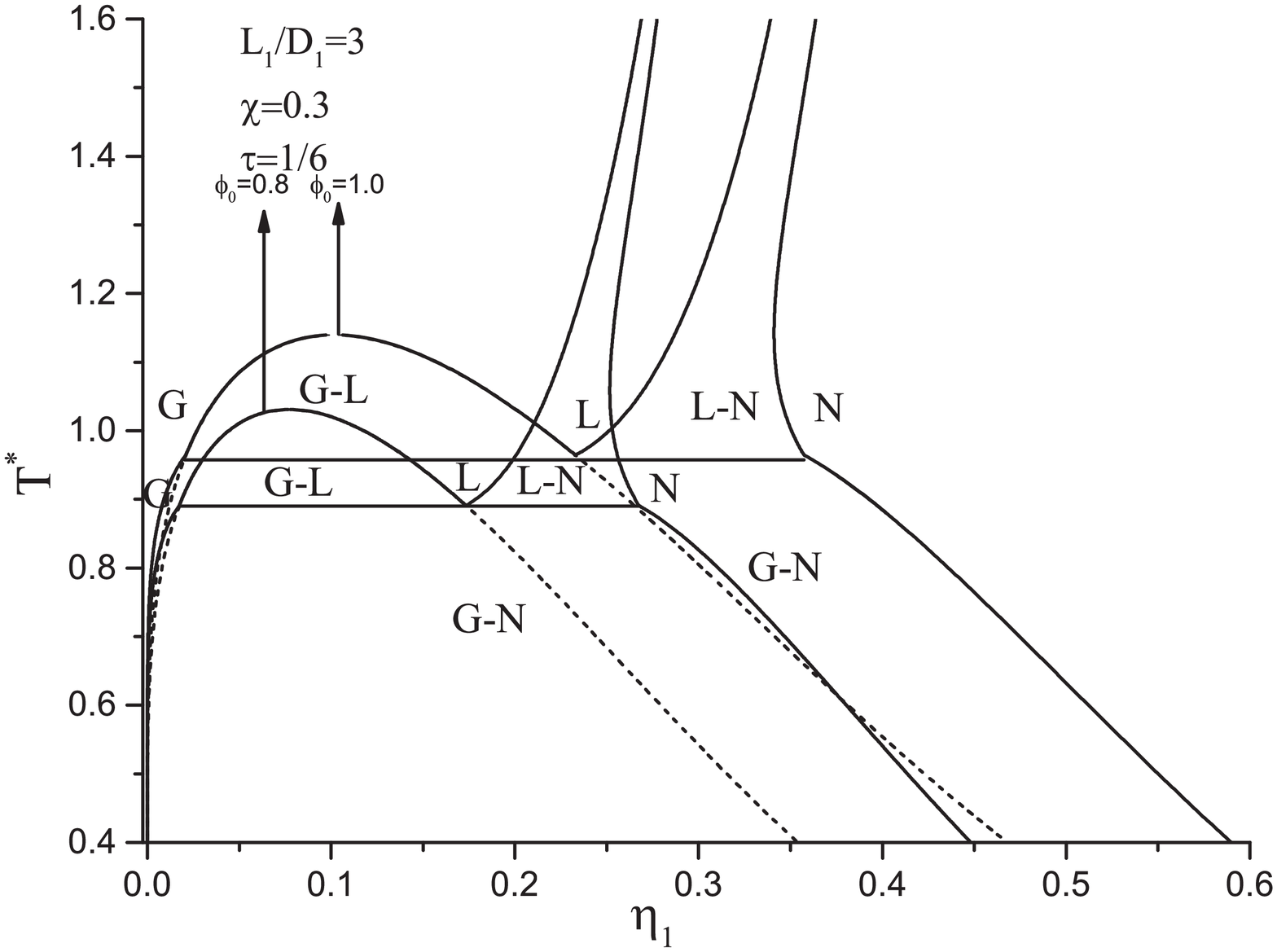}\\
{\it a}\\[3mm]
\includegraphics[width=\column]{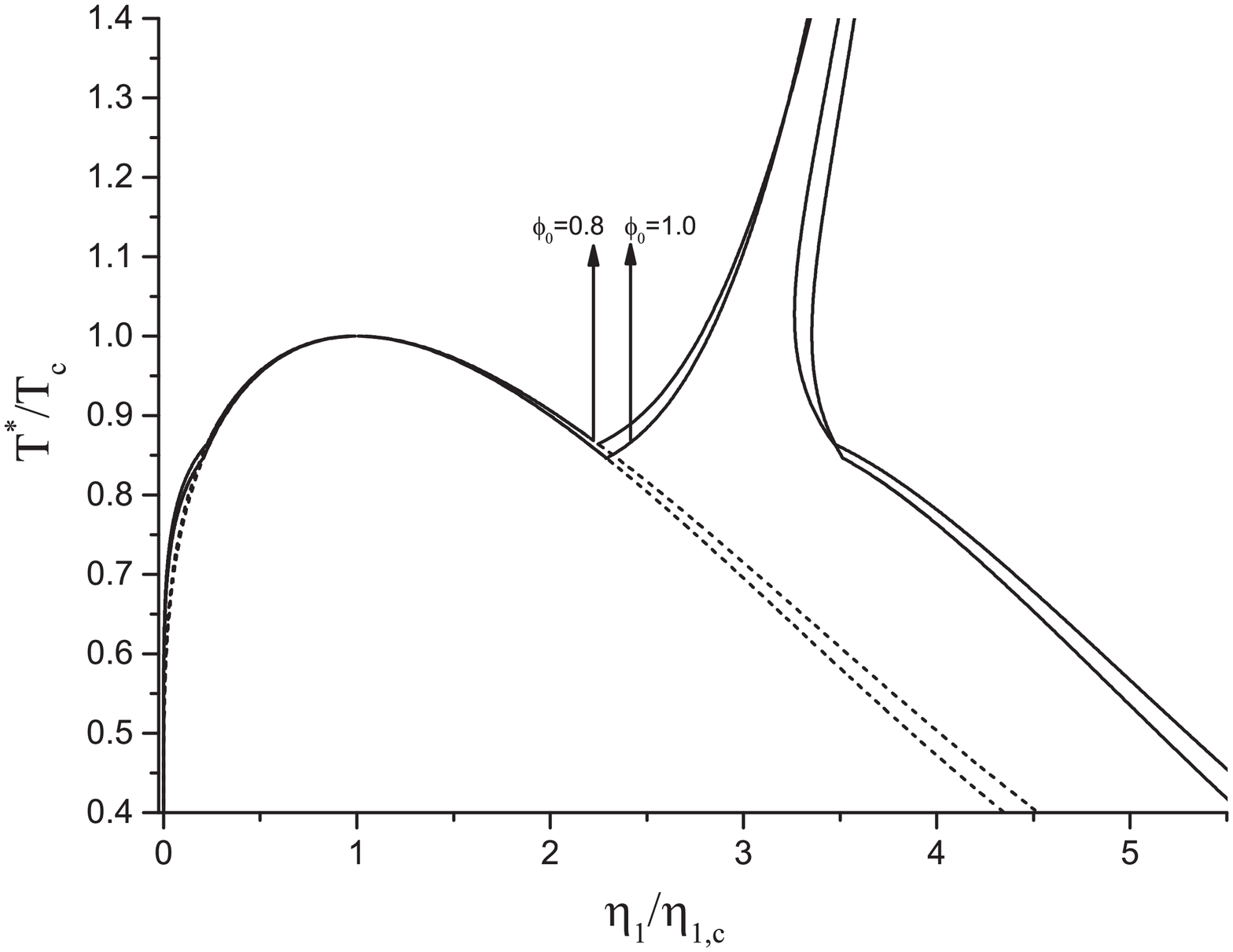}\\
{\it b} \vskip-3mm\caption{Temperature-density phase diagram
calculated from the generalized van der Waals equation for hard
spherocylinders with the anisotropy of sizes $L_{1}/D_{1}=3$ and
with isotropic and anisotropic attractive interactions ($\chi=0.3$)
in  random porous media with the porosity $\phi_{0}=0.8$ and
$\tau=1/6$.\,\,For the purpose of comparison, the phase diagram in
the absence of a porous medium $\phi_{0}=1.0$ is also
presented.\,\,The dashed lines correspond to the case
$\chi=0$.\,\,On the top (case $a$), the phase diagram in the
coordinates
 $T^{*}=kT/\epsilon_{0}$ and $\eta_{1}=\rho_{1}V_{1}$ is presented.\,\,At the bottom (case $b$),
 the phase diagram is presented in the coordinates $T^{*}/T^{*}_{c}$ and $\eta_{1}/\eta_{1,c}$  }\label{Fig2}
\end{figure}

\begin{figure}%
\vskip1mm
\includegraphics[width=\column]{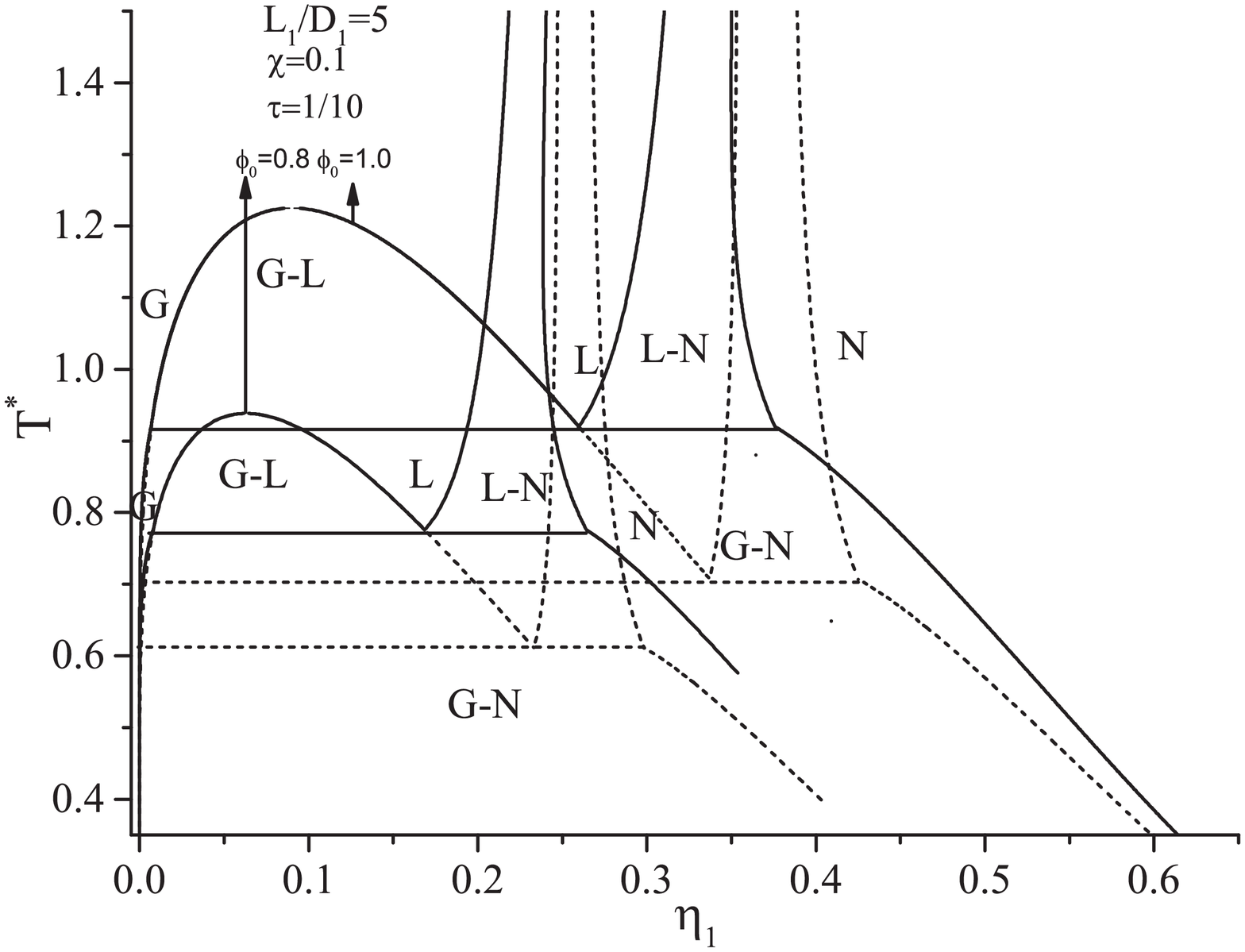}\\
{\it a}\\ [3mm]
\includegraphics[width=\column]{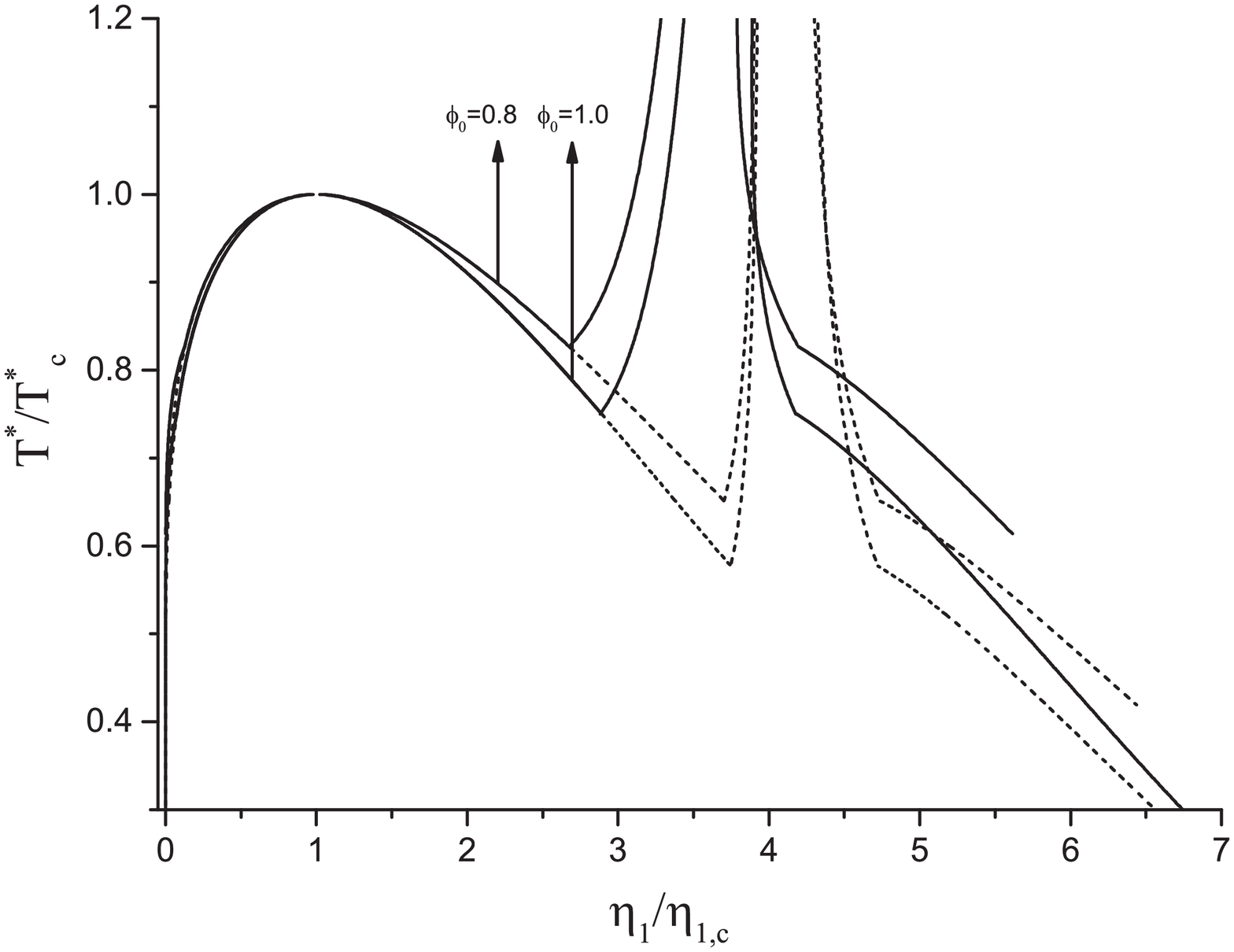}\\
{\it b} \vskip-3mm\caption{Temperature-density phase diagram
calculated from the generalized van der Waals equation for hard
spherocylinders with the anisotropy of sizes $L_{1}/D_{1}=5$ and
with isotropic and anisotropic attractive interactions ($\chi=0.3$)
in random porous media with the porosity $\phi_{0}=0.8$ and
$\tau=0.1$. For the purpose of comparison, the phase diagram in the
absence of a porous medium, $\phi_{0}=1.0,$ is also
presented.\,\,The  dashed lines correspond to the case
$\chi=0$.\,\,On top (case $a$), the phase diagram in the coordinates
 $T^{*}=kT/\epsilon_{0}$ and $\eta_{1}=\rho_{1}V_{1}$ is presented.\,\,At the bottom (case
 $b$),
 the phase diagram is presented in the coordinates $T^{*}/T^{*}_{c}$ and $\eta_{1}/\eta_{1,c}$  }\label{Fig3}
\end{figure}

\begin{figure}%
\vskip1mm
\includegraphics[width=\column]{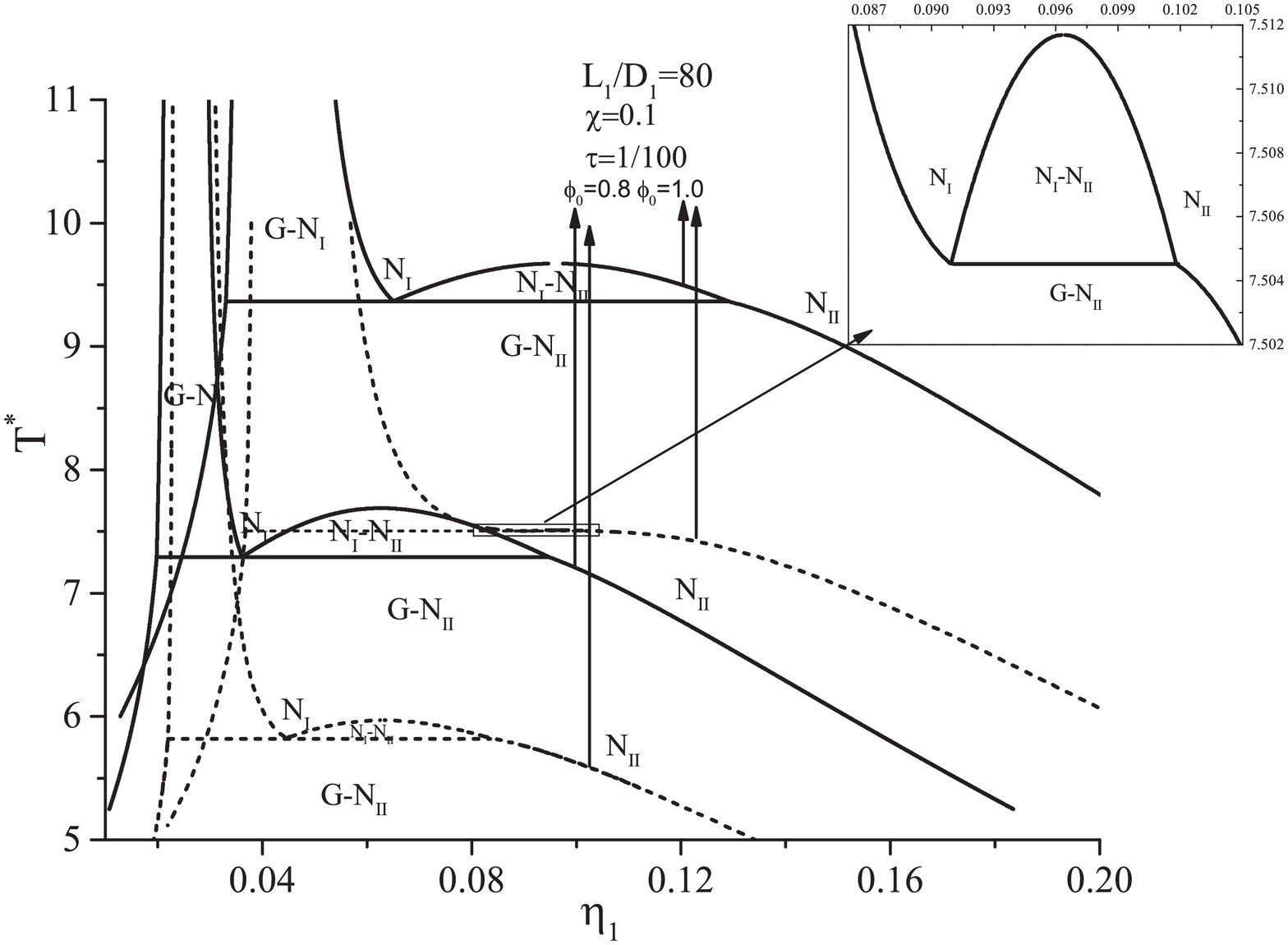}\\
{\it a}\\[3mm]
\includegraphics[width=\column]{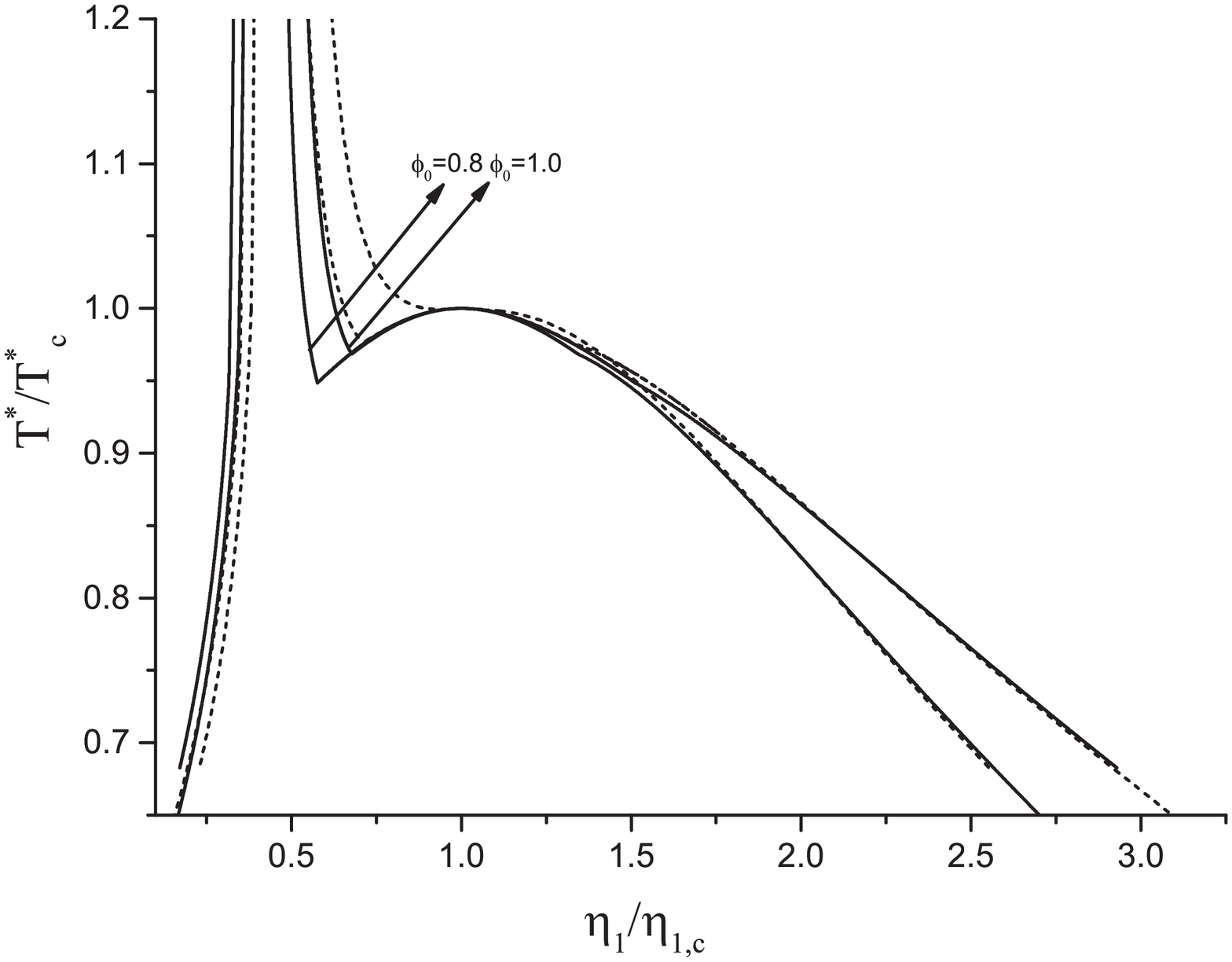}\\
{\it b} \vskip-3mm\caption{Temperature-density phase diagram
calculated from the generalized van der Waals equation for hard
spheres $L_{1}/D_{1}=80$
with isotropic and anisotropic attractive interactions ($\chi=0.3$)
in random porous media with the porosity $\phi_{0}=0.8$ and
$\tau=0.01$.\,\,For the purpose of comparison, the phase diagram in
the absence of a porous medium, $\phi_{0}=1.0,$ is also presented.
The dashed lines correspond to the case $\chi=0$.\,\,On top (case
$a$), the phase diagram in the coordinates
 $T^{*}=kT/\epsilon_{0}$ and $\eta_{1}=\rho_{1}V_{1}$ is presented.\,\,At the bottom (case $b$),
 the phase diagram is presented in the coordinates $T^{*}/T^{*}_{c}$ and $\eta_{1}/\eta_{1,c}$  }\label{Fig4}
\end{figure}

In contrast to \cite{Hol17}, where our investigations were
concentrated on the influence of the molecular shape on the phase
behavior of molecular fluids with $\chi=0$, we will focus more on
the influence of the anisotropy of attractive intermolecular
interactions.\,\,The corresponding results of our investigation of
the influence of parameter $\chi$ at different values of porosity
$\phi_{0}$ and different values of parameter $L_{1}/D_{1}$ are
presented in Figs.~\ref{Fig1}--\ref{Fig4}.\,\,The results are
presented in the form of the phase diagram in the coordinates
``dimensionless density $\eta_{1}$~-- dimensionless temperature
$T^{*}=kT/\epsilon_{0}$''.\,\,In order to compare the influence of
the parameter $\chi,$ we also present the results for
$\chi=0$.\,\,We note that, in accordance with (\ref{Hol37}), the
contribution of anisotropic attractive interactions is proportional
to
\begin{equation}
\int f(\Omega_{1})f(\Omega_{2})P_{2}(\cos\vartheta_{12})d\Omega_{1}d\Omega_{2}=S_{2}^{2},
\label{Hol44}
\end{equation}
where
\begin{equation}
S_{2}=\int P_{2}(\cos\vartheta)f(\Omega)d\Omega
\label{Hol45}
\end{equation}
is the order parameter.

Since $S_{2}=0$ in the isotropic phase, the influence of anisotropic
attractions in the isotropic phase is negligible in the van der
Waals approximation.\,\,In order to check the law of corresponding
states in each figure, we present the phase diagrams also in the
reduced variables $\eta_{1}/\eta_{1c}-T^{*}/T^{*}_{c}$, where
$\eta_{1c}$ and $T^{*}_{c}$ are the corresponding values of critical
density and critical temperature for the gas-liquid phase
\mbox{transition.}

We start from the hard sphere model with attractive interactions in
the form (\ref{Hol36}).\,\,In this case, $L=0$, and $D=\sigma$ is
the diameter of hard spheres.\,\,The corresponding phase diagram is
presented in Fig.~\ref{Fig1}.\,\,In the region of small densities,
we have the gas phase $G,$ which changes, as the density increases,
to the liquid phase $L$ and, at higher densities, to the anisotropic
nematic phase $N$.\,\,The nematic phase appears due to the
anisotropy of attractive interactions, and the bifurcation line of
the nematic phase is proportional to
$\frac{\rho_{1}\beta\chi}{\phi_{0}},$ according to (\ref{Hol41}).
This means that the phase transition appears at higher densities and
{\it vice versa}, as the temperature increases.\,\,As the
temperature decreases, the phase transition appears at lower
densities.\,\,At sufficiently low temperatures, the region of the
gas-liquid transition converges to the tricritical
gas-liquid-nematic $(G-L-N)$ point.\,\,Below the tricritical point,
only the gas-nematic $(GN)$ coexistence is seen.\,\,At temperatures
higher than the tricritical one, the anisotropic attractive
interaction does not change the gas-liquid coexistence line in the
van der Waals approximation.\,\,But, for lower temperatures, the
anisotropic attraction slightly widens the gas-liquid coexistence
and leads to a strong widening for larger densities.\,\,The presence
of a porous medium shifts the phase diagram to the region of lower
densities and lower temperature, as the porosity $\phi_{0}$
decreases.\,\,In contrast to the usual gas-liquid phase transition
\cite{Hol31}, the law of corresponding states in the considered case
is more or less valid in the region of small densities.\,\,But, at
higher densities, the coexistence curve becomes wider, as the
porosity \mbox{decreases.}\looseness=1

 The influence of the parameter $L_{1}/D_{1}$ responsible for the anisotropy of a
molecular shape is presented in Figs.~\ref{Fig2}--\ref{Fig4}.
Comparing Fig.~\ref{Fig1} and Fig.~\ref{Fig2}, we can see that the
non-sphericity of a molecular shape leads to a shift of the phase
diagram to lower densities and higher temperatures.\,\,The increase
of the ratio $L_{1}/D_{1}$ leads to an increase of the critical
temperature and a decrease of the critical density for the
gas-liquid transition.\,\,The tricritical temperature increases
also, and the gas-liquid region is essentially narrower than in the
case of
 spherical molecules.\,\,Similarly as for $L_{1}/D_{1}=0$ (Fig.~\ref{Fig1}), the anisotropic phase appears in
 the case $L_{1}/D_{1}=3$ (Fig.~\ref{Fig2}) due to
the anisotropy of attractive intermolecular interactions. However,
the anisotropy of molecule shapes highly modifies the region of the
coexistence of isotropic and nematic phases.\,\,The liquid-nematic
coexistence region becomes wider and is not very sensitive to the
temperature.\,\,Similarly as for $L_{1}/D_{1}=0$ below the
tricritical temperature, the anisotropic attractive interaction does
not change the gas-liquid coexistence line.\,\,For lower
temperatures, the coexistence region slightly widens at lower
densities and widens significantly at higher densities.\,\,We note
that, at $\chi=0,$ the nematic phase does not appear.\,\,In
agreement with the data of computer simulations \cite{Bolhuis32},
the nematic phase in the fluid of hard spherocylinders appears at
$L_{1}/D_{1}>3.7$.\,\,For $L_{1}/D_{1}=3$ similarly as for
$L_{1}/D_{1}=0,$ the decrease of the porosity $\phi_{0}$ shifts the
phase diagram to lower densities and lower temperatures.\,\,In
contrast to the case $L_{1}/D_{1}=0$, the corresponding law in the
case $L_{1}/D_{1}=3$ is more or less satisfied for all densities,
including the region of the isotropic-nematic \mbox{transition.}

The phase diagram for the case $L_{1}/D_{1}=5$ is presented in
Fig.~\ref{Fig3}.\,\,Such asymmetry of the shape of molecules is
sufficient for the nematic ordering.\,\,As we noted in \cite{Hol17},
the anisotropic attractive interaction expands the region of
orientational ordering.\,\,The
 liquid-nematic region widens significantly if $\chi$ increases.\,\,As a consequence, the gas-liquid transition
 disappears for a sufficiently large anisotropy, and only the isotropic-nematic transition takes place.\,\,Such situation for $L_{1}/D_{1}=5$ is observed at $\chi=0.3$.
 Due to this fact, we put $\chi=0.1$ in Figs.~\ref{Fig3} and \ref{Fig4}.\,\,As we can see from Fig.~\ref{Fig3}, similarly to Figs.~\ref{Fig1}
 and \ref{Fig2}, the gas-liquid coexistence line below the tricritical temperature does not depend on the anisotropic attractive interaction.
 The anisotropic attractive interaction leads to an increase of the tricritical temperature and a decrease of the tricritical density.
 As a result, the gas-liquid region is narrower, while the liquid-nematic region covers a wide range of densities.\,\,The  presence of a porous medium
 shifts the phase diagram to lower densities and lower temperatures.\,\,The corresponding law is satisfied rather well.\,\,In the region of the anisotropic
 transition, the corresponding law is satisfied separately for $\chi=0$ and for
 $\chi=0.1$.

%
%

Finally, we consider Fig.~\ref{Fig4}, where the phase diagram for
$L_{1}/D_{1}=80$ is presented.\,\,As we noted in \cite{Hol17}, the
transition into the nematic phase shifts in this case at $\chi=0$ to
the region of small densities.\,\,As a consequence, the gas-liquid
transition appears in the nematic region.\,\,In accordance with the
classification in \cite{Varga33}, the gas phase in the nematic
region is marked as $N_{I},$ and the liquid phase in the nematic
region is marked as $N_{II}$.\,\,In contrast to the previous
figures, in the case considered in Fig.~\ref{Fig4}, the influence of
an anisotropic attractive interaction is of importance for the
entire phase diagram.\,\,The anisotropic attractive interaction does
not change significantly the critical density of the gas-liquid
transition, but it changes strongly the value of critical
temperature.\,\,In contrast to the case $L_{1}/D_{1}=5,$ the
anisotropic attractive interaction in the considered case
$L_{1}/D_{1}=80$ leads to a strong increase of the tricritical
temperature and to the expansion of the gas-liquid
coexistence.\,\,The region of coexistence of the isotropic and
nematic phases is also expanded and shifts to lower densities.\,\,In
the presence of a porous medium, the phase diagram is shifted to
lower densities and lower temperatures, as the porosity $\phi_{0}$
decreases.\,\,The corresponding law is satisfied quite well for the
isotropic phase independently of the value of $\chi$.\,\,However,
the values of $\chi$ and porosity $\phi_{0}$ have a strong influence
on the phase diagram in the anisotropic region.\,\,At the critical
point, the corresponding law is satisfied, but, for higher
densities, the behavior of the phase diagram depends significantly
on the value of porosity~$\phi_{0}$.

\section{Conclusion}\label{sec_hol5}

In this paper, we have presented the generalized van der Waals
equation for anisotropic molecular fluids in porous media.\,\,This
generalization is based on analytical expressions for the equation
of state and the chemical potential of a hard spherocylinder fluid
in a random porous medium obtained in the framework of the scaled
particle theory.\,\,The second term of the generalized van der Waals
equation is the mean value of attractive intermolecular
interactions.\,\,By minimizing the free energy of the fluid, we have
obtained a nonlinear integral equation for the singlet distribution
function, which describes the orientational ordering in the system.
This ordering is connected with the non-sphericity of molecular
shapes and the anisotropy of intermolecular attractive interactions.

The investigations based on the generalized van der Waals equation
demonstrate a wide variety of gas-liquid-nematic phase behaviors in
molecular systems depending on the anisotropy of a shape of
molecules, anisotropy of attractive interparticle interactions, and
porosity of a porous medium.\,\,Here, we have focused our attention
on the influence of the anisotropy of attractive interactions, which
is the main reason for the orientational ordering in quasispherical
molecules.\,\,It is shown that the anisotropy of molecular shapes
leads to a shift of the critical point of the gas-liquid transition
to lower densities and higher temperatures and to an increase in the
tricritical temperature.\,\,The anisotropy of molecular shapes also
expands significantly the region of coexistence of the isotropic and
nematic phases.\,\,It is shown that, for a larger anisotropy, the
nematic phase can appear due to the anisotropy of molecular shapes.
In this case, the  anisotropy of attractive intermolecular
interactions expands significantly the region of coexistence between
the isotropic and nematic phases and shifts it to the region of
lower densities and higher temperatures.\,\,Finally, at a
sufficiently large anisotropy of molecular shapes, the transition
into the nematic phase takes place at very low densities.\,\,As a
result, the gas-liquid transition takes place in the nematic region.
In all the considered cases, the decrease of the porosity of a
porous medium shifts the phase diagram to the region of lower
temperatures and lower densities.

\vskip3mm

{\it The authors express their gratitude to Taras Patsahan and Ivan
Kravtsiv for useful discussions and important suggestions during the
preparation of this \mbox{paper.}}


\rezume{М.Ф.\,Головко, В.І.\,Шмотолоха} {УЗАГАЛЬНЕННЯ РІВНЯННЯ\\
ВАН-ДЕР-ВААЛЬСА НА АНІЗОТРОПНІ\\ РІДИНИ В ПОРИСТИХ СЕРЕДОВИЩАХ}
{Представлене узагальнене рівняння Ван-дер-Ваальса на анізотропні
рідини в пористих середовищах складається з двох доданків. Перший з
них базується на рівнянні стану твердих сфероциліндрів у випадковому
пористому середовищі, отриманий в рамках методу узагальнення
масштабної частинки. Другий доданок виражається через середнє
значення потенціалу притягальної міжмолекулярної взаємодії. На
основі отриманого рівняння проведено дослідження фазової поведінки
газ--рідина--нематик молекулярних систем у залежності від
анізотропії форми молекул, анізотропії притягальної взаємодії та
пористості пористого середовища. Показано, що анізотропна фаза
формується як за рахунок анізотропної притягальної взаємодії, так і
за рахунок анізотропії форми молекул. Анізотропія форми молекул
приводить до зсуву фазової діаграми в область менших густин та вищих
температур, а анізотропія притягальної взаємодії значно розширює
область співіснування ізотропної та нематичної фаз і також зсовує її
в область нижчих густин і вищих температур. Показано, що при
достатньо великій асиметрії форми молекул фазовий перехід
рідина--газ знаходиться повністю в області нематичної фази. У всіх
випадках, що розглядаються, пониження пористості пористого
середовища зсовує фазову діаграму в область нижчих густин і
температур.}
\end{document}